\documentclass[10pt]{JHEP3}

\usepackage{epsfig,multicol,bbm}
\title{A coherent understanding of low-energy nuclear recoils in liquid xenon}

\author{Peter Sorensen \\ Lawrence Livermore National Laboratory, 7000 East Ave., Livermore, CA 94550, USA\\ E-mail: \email{pfs@llnl.gov}}

\abstract{Liquid xenon detectors such as XENON10 and XENON100 obtain a significant fraction of their sensitivity to light ($\lesssim10$~GeV) particle dark matter by looking for nuclear recoils of only a few keV, just above the detector threshold.  Yet in this energy regime a correct treatment of the detector threshold and resolution remains unclear.  The energy dependence of the scintillation yield of liquid xenon for nuclear recoils also bears heavily on detector sensitivity, yet numerous measurements have not succeeded in obtaining concordant results.  In this article we show that the ratio of detected ionization to scintillation can be leveraged to constrain the scintillation yield.  We also present a rigorous treatment of liquid xenon detector threshold and energy resolution.  Notably, the effective energy resolution differs significantly from a simple Poisson distribution.  We conclude with a calculation of dark matter exclusion limits, and show that existing data from liquid xenon detectors strongly constrain recent interpretations of light dark matter.
}
\keywords{dark matter detectors, dark matter experiments}

\begin{document}
%\maketitle

\section{Introduction} \label{intro}
The recent results of an 11 live day direct search for dark matter from the XENON100 collaboration \cite{2010aprile} have brought liquid xenon detectors into sharp focus.  If a dark matter particle scatters in the 40~kg active target, the experiment hopes to detect the scintillation photons and ionisation electrons that are produced by the recoiling xenon nucleus.  This class of technology, which notably also includes the next-generation LUX experiment \cite{2010mckinsey,2009fiorucci}, has shown excellent scalability:  the 40~kg fiducial (out of 161~kg total) 11 live days exposure from XENON100 offers a similar sensitivty to the approximately $128$ live days exposure from the CDMS collaboration \cite{2010ahmed}, due almost entirely to the increased target mass.  For reference, the state of the art in liquid xenon detectors in early 2007 was a 7.2~kg fiducial target (out of 31~kg total) \cite{2007alner}.  However, there is significant disagreement regarding the sensitivity of liquid xenon detectors to light ($\lesssim10$~GeV) dark matter particles \cite{2010collar_comments,2010xenon100_comments,2010collar_commentsII}.  The argument centers on the non-linear scintillation yield of liquid xenon for nuclear recoils, $\mathcal{L}_{eff}$, and in particular on extrapolations of $\mathcal{L}_{eff}$ below the lowest energy at which measurements are available.  Since liquid xenon dark matter experiments use the primary scintillation signal to set the energy scale for particle interactions, uncertainty in $\mathcal{L}_{eff}$ leads to uncertainty in the deposited recoil energy.  Near the detector threshold of a few keV, the uncertainty is more properly a question of whether or not a low-energy nuclear recoil offers any detectable signal.  For light dark matter, this can have order-of-magnitude implications for which values of mass $m_{\chi}$ and cross section $\sigma_{\chi}$ of a hypothetical dark matter particle are consistent with the results of the experiment.

The energy dependence of $\mathcal{L}_{eff}$ has been measured by numerous groups \cite{2010manzur,2009aprile,2009lebedenko,2009sorensen,2006chepel,2005aprile}, and substantial disagreement persists $-$ particularly at nuclear recoil energies E$_{nr}\lesssim10$~keV.  The measurement is very challenging in this regime:  a 5~keV nuclear recoil is only expected to yield 1.6 (2.4) detectable scintillation photons in XENON100 (XENON10), if $\mathcal{L}_{eff}=0.09$ at that energy  \cite{2010aprile,2009angle}.  It would seem that an alternate measurement technique is needed, or failing that, an additional means to constrain the measurement.  Fortunately, such a means exists, although it has not previously been exploited for this purpose.  A particle interaction in liquid xenon creates a number $N_e$ of ionisation electrons in addition to $N_{\gamma}$ scintillation photons.  The ratio $N_{e}/N_{\gamma}$ is already used to discriminate between nuclear recoils (from neutrons, and expected from dark matter) and electron recoils (from background gamma and beta radiation).  

In Sec. \ref{nrbands} we show how $N_{e}/N_{\gamma}$ can be used to constrain $\mathcal{L}_{eff}$, by requiring consistency with measurements of $N_{e}/N_{\gamma}$ as a function of energy.  We show reasonable lower, central and upper values of $\mathcal{L}_{eff}$ in Fig \ref{fig00}.  It turns out that some extrapolation of $\mathcal{L}_{eff}$ is necessary and justified, even below the lowest measured experimental values;  this is also shown in Fig. \ref{fig00}.  A correct calculation of dark matter sensitivity requires a correct treatment of the low-energy detector efficiency, which is discussed in  Sec. \ref{etaetas1} and Sec. \ref{fiftybox} for XENON10, and in Sec. \ref{xenon100} for XENON100.  Although Poisson fluctuations are the dominant component of the energy resolution, the effective energy resolution departs significantly from the Poisson prediction.  This is discussed in Sec. \ref{detres}, and is due in large part to the fact that experiments only look for dark matter signals in the lower half of the nuclear recoil band.  In Sec. \ref{dmlimits} we apply all these results to a calculation of dark matter exclusion limits.

\section{Preliminaries} 
\subsection{Signal production and measurement in liquid xenon} \label{sigprodlxe}
This discussion is applicable to the XENON10, XENON100, Zeplin II \cite{2007alner}, Zeplin III \cite{2009lebedenko} and LUX experiments.  Numerical examples will use properties of the XENON10 detector \cite{2010aprile_xenon10}.  A particle interaction in liquid xenon creates both excited Xe$^*$ and ionised Xe$^+$ xenon atoms \cite{1978kubota}, which react with the surrounding xenon atoms to form excimers.  The excimers relax on a scale of $10^{-8}$~s with the release of scintillation photons.  This prompt scintillation light is detected by photomultiplier tubes and is referred to as the S1 signal.  The total efficiency for a single photon to produce a single photoelectron in a photomultiplier tube is $\sim0.10$ in XENON10, and varies by detector. It depends primarily on the product of the geometric light collection efficiency of the detector, and the quantum efficiency of the photomultipliers \cite{2008sorensen}.

An external electric field $E_d$ across the liquid xenon target causes a large fraction of ionisation electrons to be drifted away from an interaction site.  The electrons are extracted from the liquid xenon and accelerated through a few mm of xenon gas by a stronger (by about $\times10$) electric field E$_e$, creating a secondary scintillation signal.  This scintillation light is detected by the same photomultiplier tubes and is proportional to the number of ionisation electrons.  It is generally referred to as the S2 signal.  XENON10 measured approximately 25~S2 photoelectrons per extracted electron \cite{2009sorensen}.

Dark matter search experiments have tended to report the response of the detector to particle interactions in terms of log$_{10}(\mbox{S2/S1})$ as a function of S1 \cite{2010aprile, 2007alner,2009lebedenko,2009angle,2008angle}.  When such a plot is produced as a result of a dedicated neutron (gamma) calibration, the result is generally referred to as the nuclear (electron) recoil band.  Typical examples can be found in \cite{2010aprile,2008angle}.  For nuclear recoils, the measured quantity S1 (in units of photoelectrons) is related to the nuclear recoil energy E$_{nr}$ via 
\begin{equation}\label{eq1}
E_{nr}=\frac{1}{ \mathcal{L}_{eff}} \cdot \frac{S1}{L_y} \cdot \frac{S_e}{S_n}.
\end{equation}
For XENON10, the light yield $L_y=3.0$~photoelectrons/keVee \footnote{The suffix `ee' refers to electron recoil equivalent energy.  The suffix `r' is usually appended to indicate nuclear recoil equivalent energy, to emphasize the quenching of the electronic signal from nuclear recoils.  In this work we always mean nuclear recoil energy unless stated otherwise, and will reserve the `keVr' unit to explicitly indicate measured nuclear recoil energy.  This is discussed in detail in Sec. \ref{etaetas1}.} for 122~keV photons, and the scintillation quenching of electron and nuclear recoils due to the electric field E$_d=0.73$~kV/cm are $S_e=0.54$ \cite{2005aprile} and $S_n=0.95$ \cite{2010manzur}.  These quantities depend on the applied electric field, which is similar for XENON100 (E$_d=0.53$~kV/cm) and higher for ZEPLIN III (E$_d=3.9$~kV/cm).  This construction of $\mathcal{L}_{eff}$ in terms of S1 and the `standard candle' $L_y$ is necessary because the probability of detecting a single scintillation photon is significantly less than one.  Physically, $L_y/S_e$ represents the total detection efficiency for primary scintillation photons.

In contrast, the measured quantity S2 (also in units of photoelectrons) is related to the nuclear recoil energy E$_{nr}$ via 
\begin{equation} \label{eq1_1}
E_{nr}=\frac{S2}{\mathcal{Q}_y},
\end{equation}
where $\mathcal{Q}_y$ is the number of detected electrons per keV.  It is possible to express $\mathcal{Q}_y$ in absolute terms because single electrons are measured with high efficiency and relative ease.  Both of these quantities are reported in \cite{2010manzur}, in which the true nuclear recoil energy is known from the initial neutron energy and the scattering angle.  Note that in contrast to $\mathcal{L}_{eff}$, $\mathcal{Q}_y$ is not referred to zero electric field.  This may lead to some residual dependence on the electric field.

\subsection{Monte carlo simulation of signal production and measurement} \label{mc_description}
Because $\mathcal{Q}_y$ and $\mathcal{L}_{eff}$ govern signal production from nuclear recoils in liquid xenon, they must also dictate the shape of the nuclear recoil band.  This observation, which has not previously been used to constrain measured values of $\mathcal{Q}_y$ and $\mathcal{L}_{eff}$, is the premise of Sec. \ref{nrbands}. Given $\mathcal{Q}_y$ and $\mathcal{L}_{eff}$ as a function of nuclear recoil energy, it is relatively easy to generate a nuclear recoil data set via monte carlo simulation.  The results of this simulation will be compared with the measured XENON10 nuclear recoil band \cite{2009angle}, for various combinations of $\mathcal{Q}_y$ and $\mathcal{L}_{eff}$.  In the interest of clarity, the following description of the simulation will therefore use numerical examples from the XENON10 detector \cite{2010aprile_xenon10}.

For each simulated event with nuclear recoil energy E$_{nr}$, which was modeled in 0.25~keV steps, the simulation output is a value of S1 and S2 as would be seen by the detector.  This was accomplished as follows: given E$_{nr}$, the expected number of detected electrons N$_e$ was calculated from $\mathcal{Q}_y$ and subjected to Poisson fluctuations.  The number of S2 photoelectrons was then obtained from N$_{S2}$~=~N$_e\cdot25~$photoelectrons/electron, taking care to account for the measured width ($\sigma/\mu=0.30$) of the single electron distribution \cite{2010aprile_xenon10}.  The expected number of S1 photoelectrons N$_{S1}$ was calculated from $\mathcal{L}_{eff}$, modeled as a discrete random hit pattern on the photomultipliers, and also subjected to Poisson fluctuations \footnote{Strictly speaking, the probability for a scintillation photon to produce a photoelectron is governed by binomial statistics.  In practice, Poisson statistics are a very good approximation.}.  Both S1 and S2 were subjected to Gaussian fluctuations in the size of a single photoelectron measured by the photomultipliers, using the average $\sigma/\mu=0.58$ \cite{2010aprile_xenon10}.  Finally, the S1 signal was required to satisfy an $n\ge2$ coincidence requirement in the photomultipliers \cite{2008angle}, within a 300~ns time window \cite{2009sorensen}.  The primary scintillation signal was assumed to have a pulse shape as shown in Fig. \ref{figA}.  The scintillation decay time is discussed in the Appendix.

It is well-known that the electron and photon signals in liquid xenon are anti-correlated, for electromagnetic scattering \cite{2003conti,2007aprile}.  Anti-correlation specifically refers to the situation in which an ionisation electron recombines with a Xe$_2^+$, leading to the release of a scintillation photon.  This is described in detail by \cite{2010manzur,2005aprile} and others.  Recombination fluctuations are evident in gamma lines \cite{2009sorensen,2003conti}, and contribute significantly to the observed width of the electron recoil band \cite{2010aprile,2008angle}. Surprisingly, there is scant evidence of significant recombination fluctuations for nuclear recoils.  Recent modeling \cite{2009dahl} of signals from nuclear recoils in liquid xenon explains this as originating from an initial production ratio Xe$^*$/Xe$^+\sim0.9$ for nuclear recoils, compared with Xe$^*$/Xe$^+=0.06$ for electron recoils.  Physically this means that for nuclear recoils, much less of the recoil energy produces ionization;  this in turn significantly limits any potential effect of recombination fluctuations.  It may also explain why the nuclear recoil band falls below the electron recoil band.  The initial production ratio of Xe$^*$/Xe$^+$ is weakly dependent on E$_d$, considering the available data on scintillation quenching \cite{2010manzur,2005aprile}.  There may also be a small dependence on nuclear recoil energy, due to track structure of the recoiling nucleus.  We assume that these details are already encoded in the observed nuclear recoil band $N_{e}/N_{\gamma}$, reported as log$_{10}$(S2/S1). 

Effects arising from recombination fluctuations were deemed to be sub-dominant for nuclear recoils and were not modeled in this simulation.  If this assumption were incorrect, the simulated width of the nuclear recoil band would be too small.  As we shall see in Sec. \ref{nrbands}, the simulation is able to make a very reasonable reconstruction of the observed nuclear recoil band width. A note on boundary conditions is in order.  In all cases, we set $\mathcal{Q}_y=0$ and $\mathcal{L}_{eff}=0$ at $E_{nr}\leq0.5$~keV.  However, in all cases, the S1 $n\ge2$ coincidence requirement rejects events with such low energy.  This will be discussed in more detail in Sec. \ref{lowedet}.

\section{Nuclear recoil band constraints on $\mathcal{Q}_y$ and $\mathcal{L}_{eff}$}\label{nrbands}
We will consider three case studies to illustrate the utility and limitations of using the nuclear recoil band to constrain $\mathcal{Q}_y$ and $\mathcal{L}_{eff}$.  This is simply a statement that the energy dependence of $N_e/N_{\gamma}$, and therefore also $\mathcal{Q}_y/\mathcal{L}_{eff}$, are encoded in the reported nuclear recoil band log$_{10}$(S2/S1).  The cases are shown in Fig. \ref{fig0} as solid curves, each of which was modeled as a cubic spline.  They are truncated at the value of E$_{nr}$ below which no signal would have been recorded.  No extrapolation of $\mathcal{Q}_y$ or $\mathcal{L}_{eff}$ was made (nor would it be relevant), beyond the extent of the curves shown in Fig. \ref{fig0}. This cutoff is enforced primarily by the S1 $n\ge2$ coincidence requirement, in conjunction with the 12~electron S2 threshold.

\FIGURE{
\epsfig{file=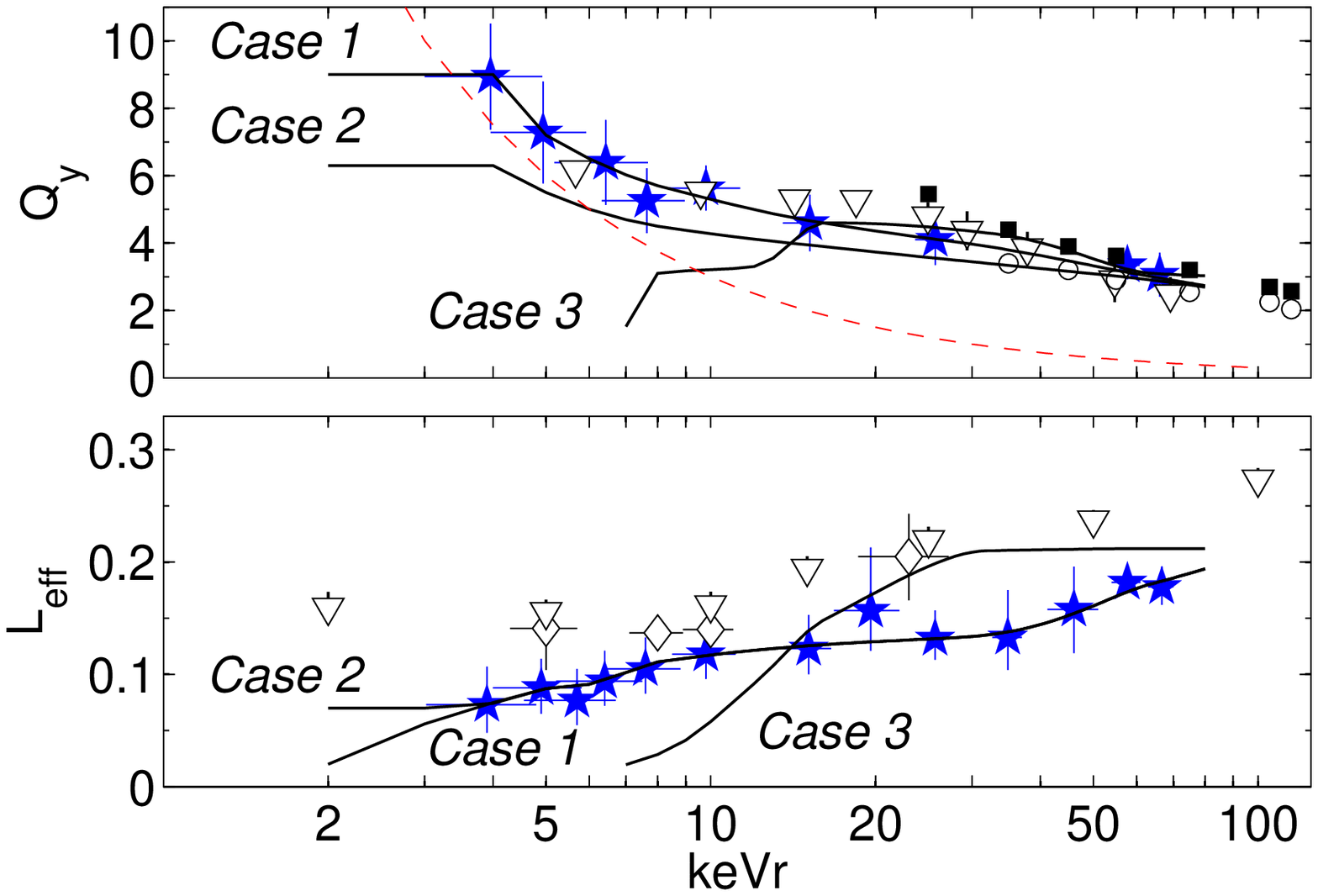,width=10cm}
\caption{$\mathcal{Q}_y$ and $\mathcal{L}_{eff}$ values used as input to the simulation (solid curves), for each case study.  Also shown are data from \cite{2010manzur} (blue $\bigstar$, E$_d=1.0$~kV/cm), \cite{2009aprile} ($\lozenge$), \cite{2009sorensen} ($\triangledown$, E$_d=0.73$~kV/cm) and \cite{2006aprile} ($\circ$, E$_d=0.1$~kV/cm and $\blacksquare$, E$_d=2.0$~kV/cm).  The top panel also shows the trend of $\mathcal{Q}_y$ corresponding a measurement of 30~electrons, regardless of nuclear recoil energy (red dash).  $\mathcal{Q}_y$ may depend on E$_e$ in addition to the weak dependence on E$_d$;  the former is often not explicitly reported. } %The data indicated by $\bigstar$ were obtained at E$_d=1.0$~kV/cm, and are $\lesssim\frac{1}{2}\sigma$ lower than measurements (not shown) obtained in the same detector at E$_d=4.0$~kV/cm.  Data indicated by $\triangledown$, $\circ$, $\blacksquare$ had E$_d=0.73,0.1,2.0$~kV/cm, respectively.}
\label{fig0}
}

\subsection{Case 1} \label{case1}
As a starting point, we take a spline through the central values of $\mathcal{Q}_y$ and $\mathcal{L}_{eff}$ from \cite{2010manzur} and check if our simulation reproduces the measured XENON10 nuclear recoil band (which we take from \cite{2009angle}).  These curves are labeled {\it Case 1} in Fig. \ref{fig0}.  A notable difference between the Manzur {\it et al.}, \cite{2010manzur} experiment and \cite{2009angle} is the different electric field $E_d$ across the active target.  While the difference between $E_d=1$~kV/cm and $E_d=0.73$~kV/cm appears to be negligible for scintillation quenching of nuclear recoils \cite{2005aprile}, it may have a small effect on $\mathcal{Q}_y$.  We will discuss this further in Sec. \ref{case2}.

\FIGURE{
\epsfig{file=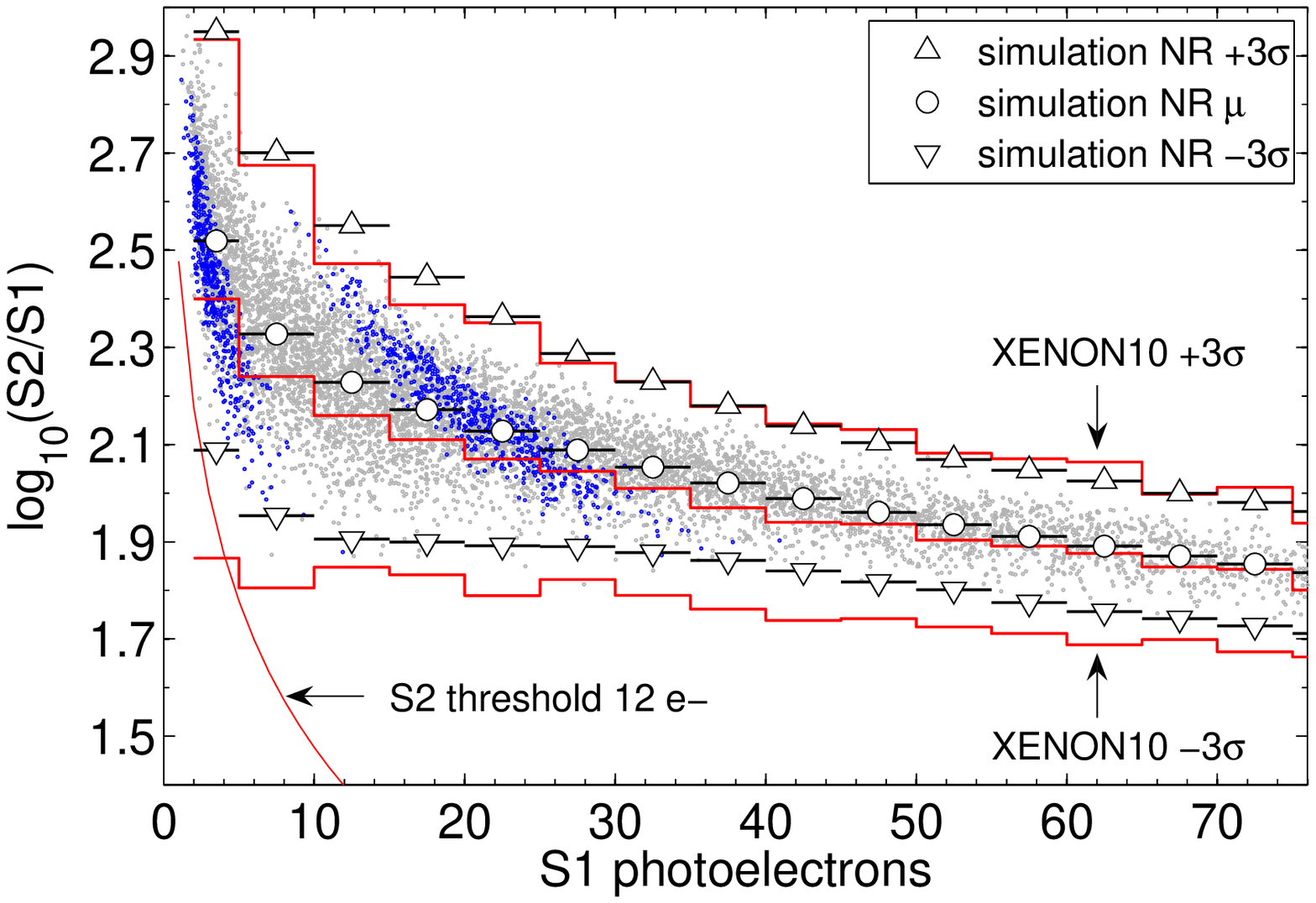,width=10cm}
\caption{The XENON10 nuclear recoil band parameterization centroid and $\pm3\sigma$ (red stair steps) \cite{2009angle} are compared with the simulated nuclear recoil band data (light grey points).  Simulated data corresponding to E$_{nr}=5$~keV and E$_{nr}=30$~keV are highlighted in dark blue to indicate the mapping between the nuclear recoil band and true nuclear recoil energy.  Resulting centroid and $\pm3\sigma$ contours from Gaussian fits to the simulated data are indicated in the figure legend.  The S2 software threshold that XENON10 implement is duplicated in the simulation.  }
\label{fig1}
}

Figure \ref{fig1} shows the XENON10 nuclear recoil band $\mu$ and $\pm3\sigma$ contours (red stair steps), along with the 12~electron S2 threshold (solid red curve) described in \cite{2009angle}.  The simulated data points are shown in light gray, and data points corresponding to E$_{nr}=5$~keV and E$_{nr}=30$~keV are highlighted in dark blue to elucidate the mapping between the nuclear recoil band and true nuclear recoil energy.  The XENON10 collaboration report that their nuclear recoil band was calculated in discreet S1 bins, by fitting a Gaussian to the log$_{10}(\mbox{S2/S1})$ distribution \cite{2009angle}.  This distribution is empirically highly Gaussian \cite{2010aprile_xenon10}, with some non-Gaussian tails evident on the low side.  

We employed exactly the same procedure and bin widths in calculating the nuclear recoil band obtained from our simulation, and display the $\mu$ (as $\circ$) and $\pm3\sigma$  (as $\vartriangle$ and $\triangledown$) values, with horizontal bars indicating the bin width.  Simulated events which appear below the 12 electron threshold were discounted prior to making a Gaussian fit.  The statistical error in the simulated $\mu$ and $\pm3\sigma$ values, if not visible, is smaller than the size of the data points.  Bearing in mind that the $y$ axis in Fig. \ref{fig1} is logarithmic in S2/S1, the systematic disagreement between our simulated $\mu$ and the XENON10 measured $\mu$ is severe.  Evidently, self-consistency requires a decrease in $\mathcal{Q}_y$, or an increase in $\mathcal{L}_{eff}$.

Although it is not discussed in detail here, we also performed this same analysis using a spline through the central $\mathcal{Q}_y$ and $\mathcal{L}_{eff}$ values reported by XENON10 \cite{2009sorensen}.  These values are indicated by $\triangledown$ in Fig. \ref{fig0}.  The simulated $\mu$ were systematically lower than the measured XENON10 $\mu$, indicating that either an increase in the measured $\mathcal{Q}_y$, or a decrease in $\mathcal{L}_{eff}$ would be necessary for \cite{2009sorensen} to obtain self-consistency.  An example of the requisite decrease in $\mathcal{L}_{eff}$ that would yield self-consistency is provided by the dotted curve in Fig. \ref{fig00}.

\subsection{Case 2} \label{case2}
The $\mathcal{Q}_y$ indicated by the curve labeled {\it Case 2} in Fig. \ref{fig0} was constructed to obtain consistency between the central values of $\mathcal{L}_{eff}$ reported in \cite{2010manzur}, the XENON10 nuclear recoil band and the simulated nuclear recoil band.  The agreement is quite good, as shown in Fig. \ref{fig2}.  Note that this $\mathcal{Q}_y$ curve is better than $1\sigma$ consistent with the measurement of \cite{2010manzur}, except below about 8~keV.  The fact that the XENON10 data in this case require a systematically lower $\mathcal{Q}_y$ at all energies may simply be a reflection of the differences in $E_d$, as discussed in Sec. \ref{case1}.  The flat extrapolation of $\mathcal{L}_{eff}=0.07$ between 1.2~keV and 4~keV was necessary to align the $\mu$ values of the nuclear recoil band, in the $2-5$~S1 photoelectron bin.  The falling extrapolation of $\mathcal{L}_{eff}$ as illustrated by {\it Case 1} is perhaps more aligned with theoretical expectations \cite{1963lindhard}.  However, agreement with the XENON10 nuclear recoil band in that case would have required $\mathcal{Q}_y$ to be $2\sigma$ below the measured values \cite{2010manzur} at 4~keV and 5~keV.  In general one should expect a measurement of $\mathcal{Q}_y$ to be significantly more robust than a measurement of $\mathcal{L}_{eff}$.  For example, a typical 6~keV nuclear recoil in the XENON10 detector has the expectation (from Fig. \ref{fig2}) of 3~S1 photoelectrons and about 750~S2 photoelectrons (30 ionisation electrons).  In contrast to the S1 signal, the S2 signal is (in general) far above threshold and far from the regime in which Poisson fluctuations dominate.  These considerations imply that a $\mathcal{Q}_y$ curve lower than indicated by {\it Case 2} would be difficult to explain.  As it stands, there remains some tension with the measured value at 4~keV \cite{2010manzur}.  This would argue for higher values of $\mathcal{L}_{eff}$ below 8~keV. 

\FIGURE{
\epsfig{file=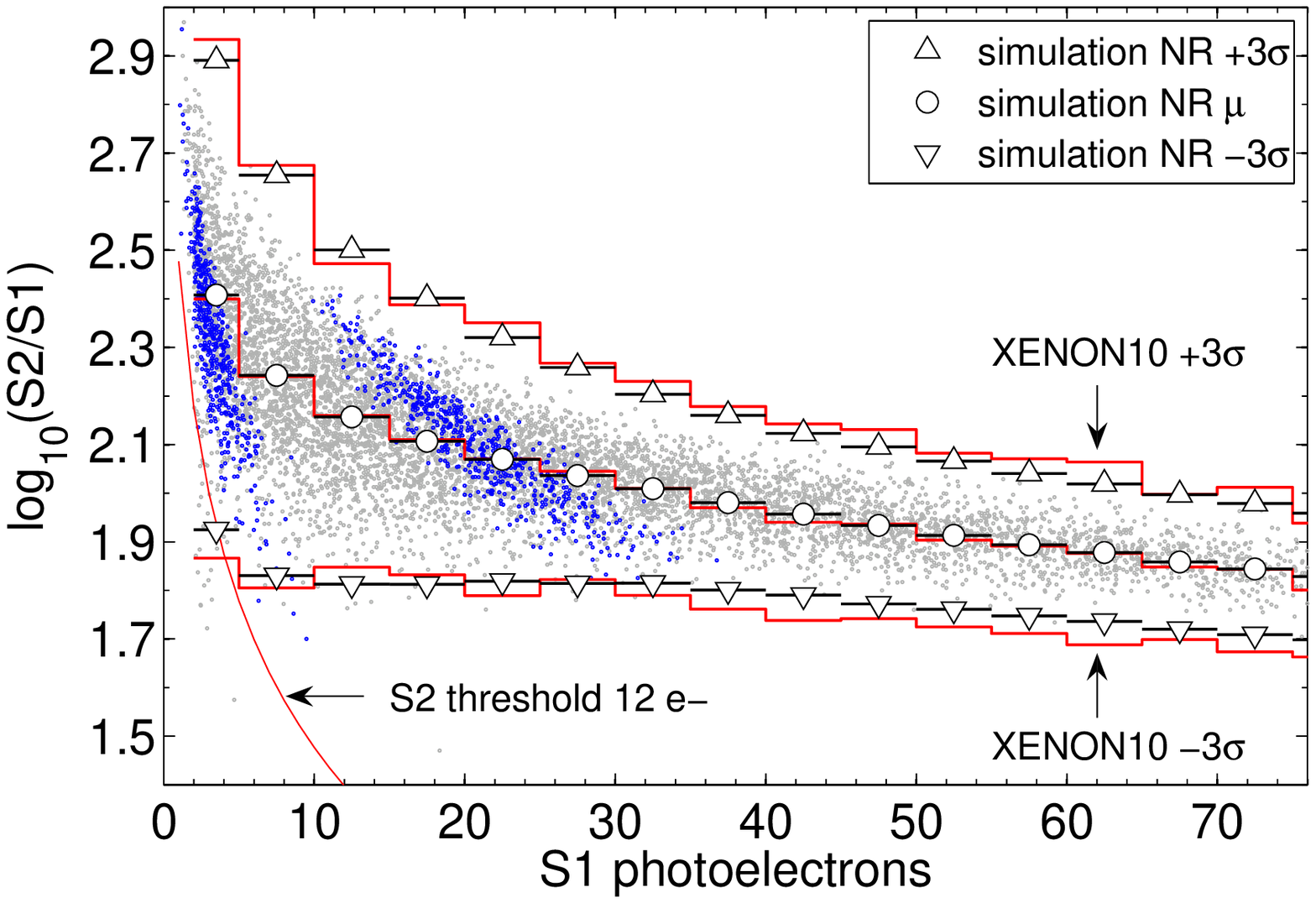,width=10cm}
\caption{All symbols and curves as described in Fig. \ref{fig1}.}
\label{fig2}
}

A possible resolution lies in the S2 trigger threshold.  The detector used in \cite{2010manzur} appears to have an S2 gain of about 8~photoelectrons per ionisation electron (about $\times3$ lower than XENON10), based on the mean value of $\sim300$~S2 photoelectrons measured for E$_{nr}\simeq6$~keV (see \cite{2010manzur}, Fig. 8c).  Suppose that the S2 trigger efficiency begins to drop below about 180~S2 photoelectrons (as it does in \cite{2010aprile_xenon10}). 
This would cut off the low side of the measured S2 distribution, and thereby artificially inflate the measured central S2 value at low recoil energies.  Figure \ref{fig0} shows the $\mathcal{Q}_y$ curve that would result if \cite{2010manzur} measured a central value of 240~S2 photoelectrons (30~electrons) at all recoil energies (red dash curve).  Above E$_{nr}\simeq8$~keV this is clearly unrealistic, but for the lowest nuclear recoil energies it could indicate a measurement trend imposed by the S2 trigger efficiency of the detector.  This could ease the tension between the measured value of $\mathcal{Q}_y$ at 4~keV, and the $\mathcal{Q}_y$ curve required for simultaneous consistency with the central $\mathcal{L}_{eff}$ values of \cite{2010manzur} and the XENON10 nuclear recoil band.

\subsection{Case 3} \label{case3}
In order for the Zeplin III measurement of $\mathcal{L}_{eff}$ to achieve consistency with the XENON10 nuclear recoil band, an abrupt drop in $\mathcal{Q}_y$ is required.  The level of consistency is not shown, but appears similar to that shown in Fig. \ref{fig2} (without the dark blue band corresponding to 5~keV events).  The requisite crash in $\mathcal{Q}_y$, labeled {\it Case 3} in Fig. \ref{fig0}, cannot be reconciled with measurements \cite{2010manzur}.  Given the robustness of the S2 signal, we are forced to consider other explanations.  One possibility is that the electron extraction efficiency in the Zeplin III detector is significantly lower than in XENON10.  This effect is expected, although there is some uncertainty in its magnitude, because the electric field across the liquid-gas interface is about 25\% lower in Zeplin III than in XENON10.  However, such an effect would be expected to result in a uniform systematic decrease of $\mathcal{Q}_y$ as a function of recoil energy.  Other possibilities are that the Zeplin III S1 detection efficiency may be lower than stated in Fig. 14 of \cite{2009lebedenko}.

\subsection{Summary: most reasonable estimate of $\mathcal{Q}_y$ and $\mathcal{L}_{eff}$}
At this point we pause to make a few observations.  First, the $\mathcal{Q}_y$ and $\mathcal{L}_{eff}$ curves labeled {\it Case 2} in Fig. \ref{fig0} are reasonably consistent with recent measurements \cite{2010manzur,2009aprile,2006aprile}, and the simulation indicates consistency with the XENON10 nuclear recoil band $\mu$ and $\sigma$.  The reproduction of the observed XENON10 band width $\sigma$ confirms our assumption that recombination fluctuations are subdominant.  The {\it Case 2} $\mathcal{Q}_y$ curve is as low as can be comfortably explained by a possible systematic effect between the slightly different electric fields $E_d$, given the near absence of scintillation quenching for nuclear recoils \cite{2005aprile}.  It is important to recognize the most glaring weakness of this analysis: the lack of a complete set of $\mathcal{Q}_y$, $\mathcal{L}_{eff}$ and nuclear recoil band measurements under the same conditions.  This inevitably leads to systematic uncertainty.  As mentioned in Sec. \ref{case1}, the magnitude of this uncertainty does appear to be small.  We therefore  propose that the central values of \cite{2010manzur}, with the extrapolation indicated by {\it Case 2}, represent the most reasonable estimate of $\mathcal{L}_{eff}$ given the available data.

\subsection{A proposed calibration strategy}
Because of the tendency of liquid xenon detector experiments to report their results with an S1-based energy scale, an enormous amount of pressure has been placed on measuring $\mathcal{L}_{eff}$.  This has shifted the experimental focus away from what should be a far easier measurement, that of $\mathcal{Q}_y$.  The results of Sec. \ref{nrbands} suggest a clear course of action for future tagged neutron scattering measurements of $\mathcal{L}_{eff}$, as there will undoubtably be.  In addition to the methods and techniques already employed \cite{2010manzur,2009aprile} an additional first step should be the acquisition of nuclear recoil data with a broad-spectrum neutron source such as americium-beryllium, or californium.  This will provide a nuclear recoil band and an essential cross-check.  The second step should be to make a robust measurement of $\mathcal{Q}_y$ using a tagged neutron source of known energy.  In order to minimize the measured uncertainty on $\mathcal{Q}_y$, the gain of the S2 signal should be high, $\gtrsim25$~photoelectrons per ionisation electron. Ideally the S2 trigger threshold would be set at the level of a single ionisation electron, as XENON10 have proven possible \cite{2006sorensen}.  If that were to lead to an excessive trigger rate, a careful assessment of the trigger threshold with a voltage step input \cite{2010aprile_xenon10} could be used instead.  This, combined with the analysis detailed in Sec. \ref{nrbands}, should allow a significant reduction in systematic and statistical uncertainty in the energy dependence of $\mathcal{L}_{eff}$.

\section{Low-energy nuclear recoils} \label{lowedet}
The discussion in Sec. \ref{lowedet} assumes $\mathcal{Q}_y$ and $\mathcal{L}_{eff}$ indicated by {\it Case 2} in Fig. \ref{fig0}, and will continue to assume instrumental parameters corresponding to the XENON10 detector.

\subsection{Detector efficiency} \label{etaetas1}
Event detection efficiency $\eta$ as E$_{nr}$ approaches zero is dominated the S1 detection efficiency.  An S2 software threshold at the level of 12~electrons (XENON10) or 15~electrons (XENON100) plays a very minor role.  Our simulation returns a direct measurement of $\eta$, which is shown in Fig. \ref{fig7} ($\triangledown$) as a function of keV. 

\FIGURE{
\epsfig{file=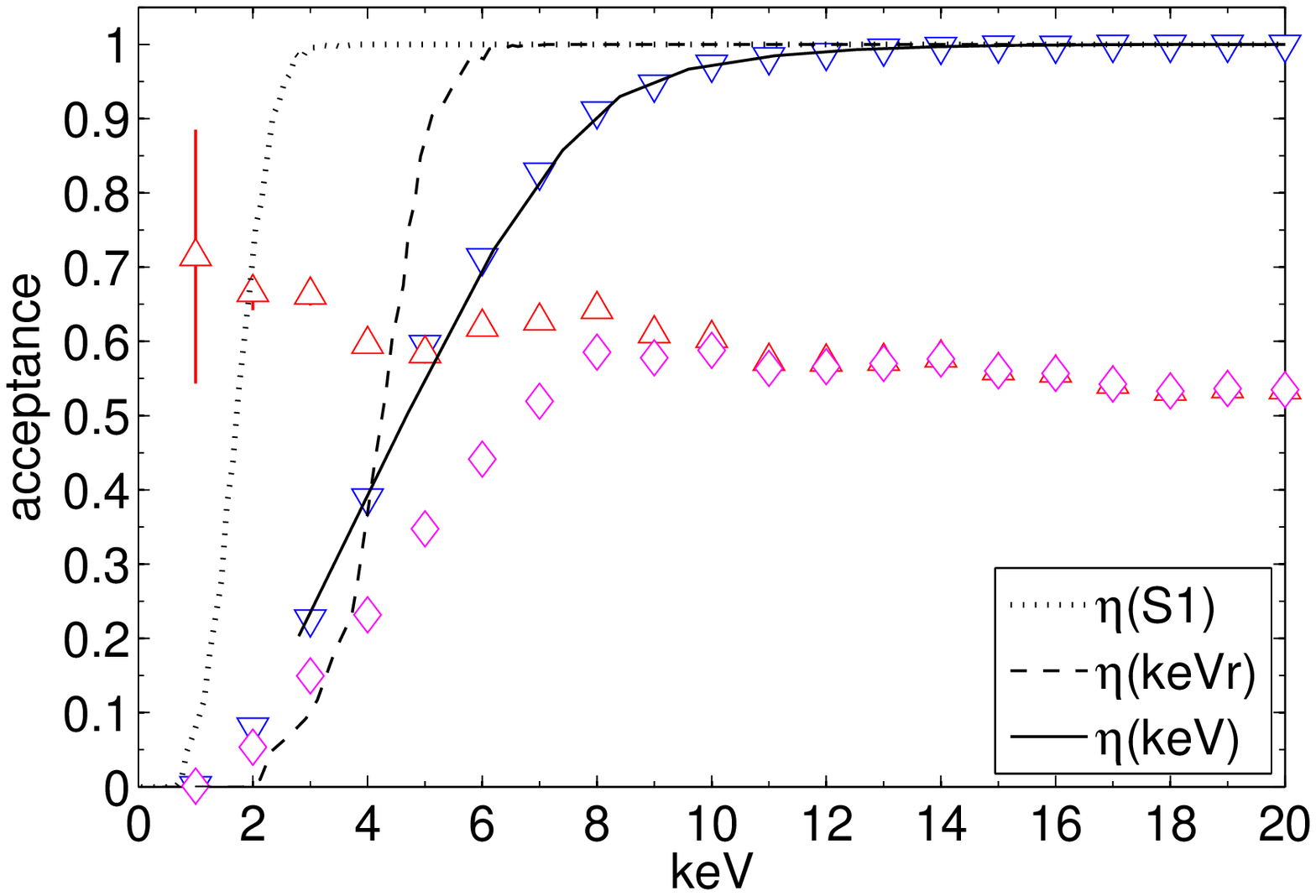,width=10cm}
\caption{The detection efficiency $\eta$(S1) (dotted curve), $\eta$(keVr) (dashed curve) obtained from Eq. \ref{eq1}, $\eta$(keV) (solid curve) obtained from Eq. \ref{fig2} and $\eta$ as obtained directly from the simulation ($\triangledown$).  Also shown are the acceptance of the `50\% acceptance box' ($\vartriangle$) and the combined low-energy acceptance for nuclear recoils ($\lozenge$).  The $x$ axis is in keV unless indicated otherwise in the legend.  Statistical uncertainty, where not visible, is smaller than the data point.}
\label{fig7}
}

A more accessible quantity is frequently reported, the behavior of $\eta$(S1)  \cite{2010manzur,2009aprile,2009sorensen,2010aprile_xenon10}. It can be obtained by a simple photoelectron statistics monte carlo simulation, and is a function of the number of detected S1 photoelectrons.  Comparison with low-energy Compton scattering background (e.g. Fig. 4.8 of \cite{2008sorensen}) can be used to confirm its validity.  In order to correctly apply the detector efficiency $\eta$ to the application of dark matter exclusion limits, it is necessary to obtain either $\eta$(keV) or $\eta$(keVr).  The distinction is important here.  As previously mentioned,  keVr is already widely used to mean `keV nuclear recoil equivalent energy,' to emphasize the quenching of the electronic signal from nuclear recoils.  The keVr unit carries the tacit assumption that it is a measured quantity.  It must therefore also carry the imprint of the detector resolution.  In contrast, the keV unit is most apt for a calculated or theoretical quantity, with perfect resolution.  There are two correct paths forward:
\begin{enumerate}
\item{convert $\eta$(S1) to $\eta$(keVr) via $\mathcal{L}_{eff}$.  The result should only be applied to a theoretical prediction of the differential scattering rate, such as Eq. \ref{rate_eq}, after first correcting for the detector resolution.}
\item{convert $\eta$(S1) to $\eta$(keV) via the method described in Sec. \ref{etakeV}.  The result can then be applied directly to a theoretical prediction of the differential scattering rate, since account has been taken of detector resolution.}
\end{enumerate}

\subsubsection{How to obtain $\eta$(keVr)} \label{etakeVr}
In Fig. \ref{fig7} we show $\eta$(S1) for {\it Case 2} (dotted curve).  It agrees quite well with $\eta$(S1) as reported \footnote{Note that in Fig. 5 of \cite{2009sorensen}, $\eta(S1)$ is referred to as `S1 peak-finding,' and is shown with a factor $\times3$ in the $x$ axis.} by the XENON10 collaboration \cite{2009sorensen}.  As an example, consider that $\eta$(2.1)$=0.8$.    From Eq. \ref{eq1} and the $\mathcal{L}_{eff}$ curve represented by {\it Case 2}, we obtain from  the conversion 2.1~S1 photoelectrons = 4.7~keVr.  Thus, $\eta(4.7~\mbox{keVr})=0.8$.  Applying this procedure to the entire $\eta$(S1) curve gives $\eta$(keVr), as shown in Fig. \ref{fig7} (dashed curve).  While both $\eta$(S1) and $\eta$(keVr) are correct representations of the detector efficiency, neither can be applied directly to a theoretical prediction of the differential dark matter scattering rate (such as Eq. \ref{rate_eq}).  The difference between these two curves and $\eta$(keV) lies in the detector resolution.

\subsubsection{How to obtain $\eta$(keV)} \label{etakeV}
To be explicit, $\mathcal{L}_{eff}$ is a measure of the expected number of detected S1 photoelectrons, given a nuclear recoil energy E$_{nr}$ (in keV).    In calculating the nuclear recoil energy E$_{nr}$, given some number of detected S1 photoelectrons, the correct response is to specify a value of E$_{nr}$ and a probability that it resulted in the observed S1 measurement.  As an example consider a measurement of 2~S1~photoelectrons.  Such an event is the expected result for an initial E$_{nr}=4.6$~keV, but E$_{nr}=8$~keV still has a 10\% chance of resulting in S1~$=2$~photoelectrons.  The probabilities are governed by $\mathcal{L}_{eff}$ and Poisson fluctuations.  To correctly calculate $\eta(4.6~\mbox{keV})$, we must account for all possibilities.  We do this by piece-wise multiplication of $\eta$(S1) with the Poisson probabilities that a measurement returned $[0~1~2~3~...]$~photoelectrons, given the expectation $k=2$~photoelectrons.  In other words, define
\begin{equation}\label{eq2}
\eta_f(k) = f_{Poisson}([0~1~2~3~...];k)
\end{equation}
for integer values of $k$ detected photoelectrons.  For $k=2$, 
\begin{equation}
f_{Poisson}([0~1~2~3~...];2) = [0.14~0.27~0.27~0.18~...].
\end{equation}
The total detection efficiency is given by
\begin{equation}\label{eq3}
\eta_{keV}(k) = \sum_i{ \eta_f(i;k) \cdot \eta_{S1}(i) }
\end{equation}
in which the nuclear recoil energy E$_{nr}$ is obtained for each $k$ via $\mathcal{L}_{eff}$.  For $k=2$, we obtain $\eta$(4.7~keV)$=0.5$. The predicted $\eta$ obtained from Eq. \ref{eq3} is shown in Fig. \ref{fig7} (solid curve), and agrees well with $\eta$ obtained directly from the simulation ($\triangledown$).  The domain of each $\eta$ appearing in Eq. \ref{eq3} is explicitly subscripted for clarity. Note that $\eta$(keV) is only defined for integer values of $k$ detected photoelectrons.   The solid curve in Fig. \ref{fig7} truncates at E$_{nr}\simeq2.8$~keV, which corresponds to $k=1$ S1 photoelectron.

\subsection{The 50\% acceptance box} \label{fiftybox}
Experiments \cite{2010aprile,2009angle,2008angle} have tended to use the region of the nuclear recoil band between $\mu$ and $\mu-3\sigma$ as the dark matter ``50\% acceptance box.''  Considering the tendency of experiments to report results as shown in Fig. \ref{fig2}, one should expect 50\% acceptance as a function of S1. In Fig. \ref{fig7} we show the acceptance of the 50\% acceptance box for nuclear recoils ($\vartriangle$) as a function of keV.  For E$_{nr}\gtrsim20$~keV, the 50\% acceptance box lives up to its name;  as E$_{nr}$ approaches zero, the acceptance of the region between $\mu$ and $\mu-3\sigma$ climbs by up to $30\%$.  This should not be surprising:  at small nuclear recoil energies, the only events which experiments are able to record have benefitted from upward Poisson fluctuations in the number of detected S1 photoelectrons.  Assuming normally distributed S2 values \footnote{This is reasonable; from Fig. \ref{fig2}, even at E$_{nr}=2$~keV, one expects to detect $\sim12$ electrons.}, such events are forced to lower values of log$_{10}(\mbox{S2/S1})$, and so in a typical nuclear recoil band plot (as a function of S1), appear preferentially below the measured band centroid.  The acceptance ($\vartriangle$) shows more scatter than might be expected from the stated uncertainty;  this is simply an effect of the discreet S1 bins employed in the calculation.  The combined low-energy acceptance of the 50\% acceptance box and $\eta$ is also shown in Fig. \ref{fig7} ($\lozenge$).

Although the nuclear recoil band as a function of detected S1 photoelectrons is empirically highly Gaussian, there is a small non-Gaussian component at low recoil energies. This non-Gaussian component is a natural result of Poisson fluctuations, although other factors that our simulation does not consider may also contribute.

\subsection{Detector resolution} \label{detres}
As mentioned, experiments have tended to report event energies in terms of detected S1 photoelectrons \cite{2010aprile,2009angle,2008angle}.  One could reasonably expect the detector resolution for nuclear recoils, as a function of E$_{nr}$, to be governed by Poisson statistics for the number of detected S1 photoelectrons.  Naturally, reality is more complicated.  

\FIGURE{
\epsfig{file=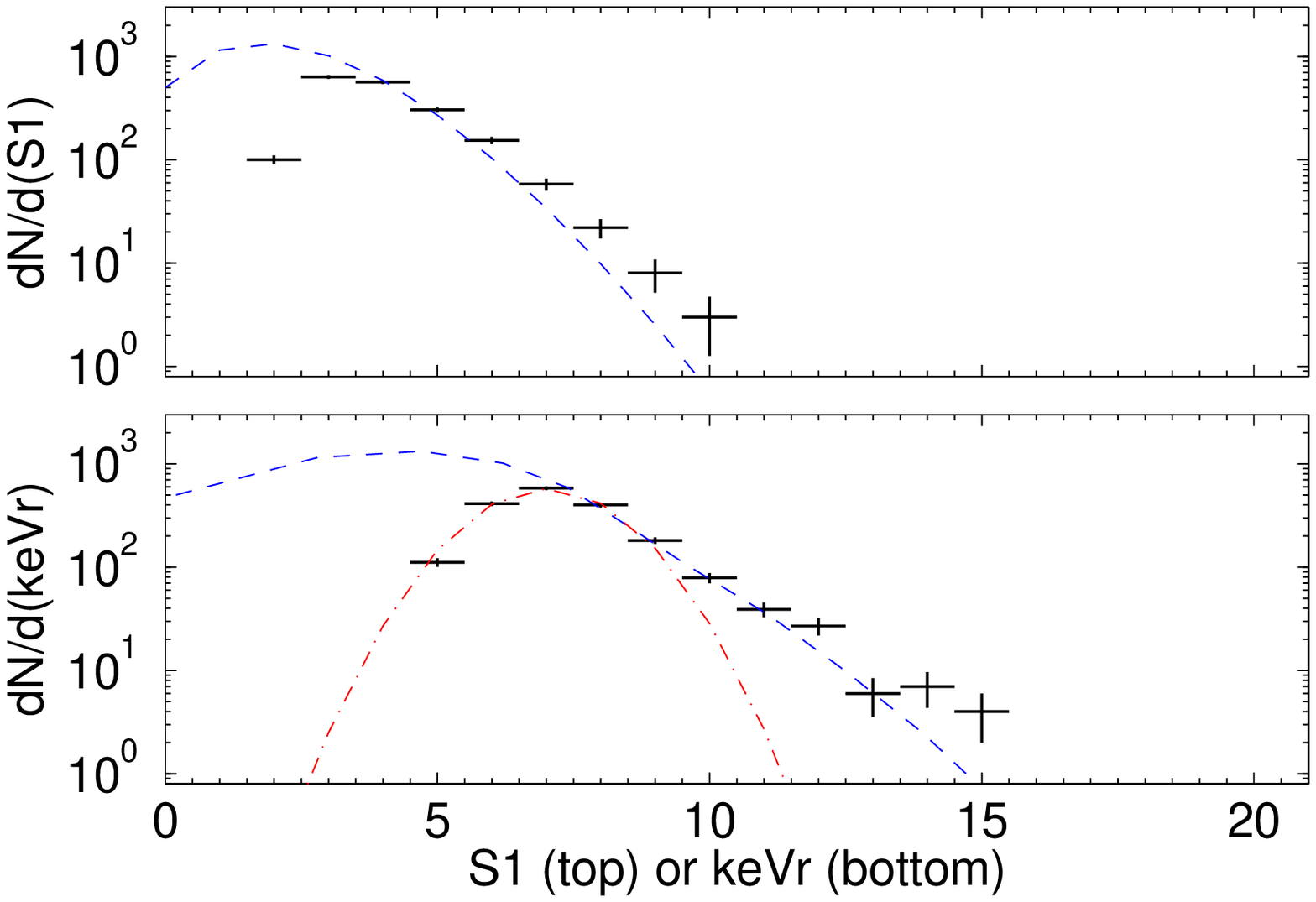,width=10cm}
\caption{{\bf (top)} S1 distribution for E$_{nr}$=5~keV, and Poisson distribution (blue dash) for the expected number ($k=2.3$) of S1 photoelectrons at this recoil energy. {\bf (bottom)} The same S1 distribution, and Poisson distribution for $k=2.3$, both transformed to keVr via $\mathcal{L}_{eff}$ from Fig. \ref{fig2}.  A Gaussian fit (red dash dot) is also shown.}
\label{fig8}
}

The simulation allows us to directly measure the detector resolution, considering the inputs listed in Sec. \ref{mc_description}.  The distribution of S1 photoelectrons for events falling in the 50\% acceptance box is shown in Fig. \ref{fig8} for the case E$_{nr}=5$~keV.  Considering {\it Case 2} shown in Fig. \ref{fig0}, XENON10 should expect to measure 2.3~S1 photoelectrons at E$_{nr}=5$~keV.  A Poisson distribution with $k=2.3$ is shown (top panel, blue dash), normalized to the total N~$=5000$ simulated events.  The same distribution is shown (bottom panel, blue dash) transposed from S1 photoelectrons to keVr using $\mathcal{L}_{eff}$, and also fails to match the simulation result.  A total of 1887 events comprise the histograms in Fig. \ref{fig8}.  This is reflected in the total acceptance ($\lozenge$) of 0.37 shown in Fig. \ref{fig7}.  An additional 1158 events with E$_{nr}=5$~keV fell in the region between $\mu$ and $\mu+3\sigma$, which is reflected in the acceptance ($\vartriangle$) of 0.61 shown in Fig. \ref{fig7}.  The remaining 1955 events either failed to satisfy the $n\ge2$ S1 coincidence requirement, or the S2~$>300$ photoelectron requirement.  This is reflected in the acceptance ($\triangledown$) of 0.61 shown in Fig. \ref{fig7}.

In terms of either S1 photoelectrons or keVr, the observed distribution is at best quasi Poisson.  The reason for this is two-fold: the left edge of the distribution is truncated by both the S1 detection efficiency and the top of the 50\% acceptance box.  Both edges are broadened by photomultiplier fluctuations.  Figure \ref{fig8} illustrates that at low recoil energies, a detector will retain sensitivity to upward fluctuations in measuring the primary scintillation signal.  However, it also illustrates that one must be explicit about the underlying assumptions when ``taking into account an S1 resolution dominated by Poisson fluctuations'' \cite{2010aprile}.  For reference Fig. \ref{fig8} (bottom) also shows a Gaussian fit (red dash-dot), although the quality of the fit is rather poor.

\FIGURE{
\epsfig{file=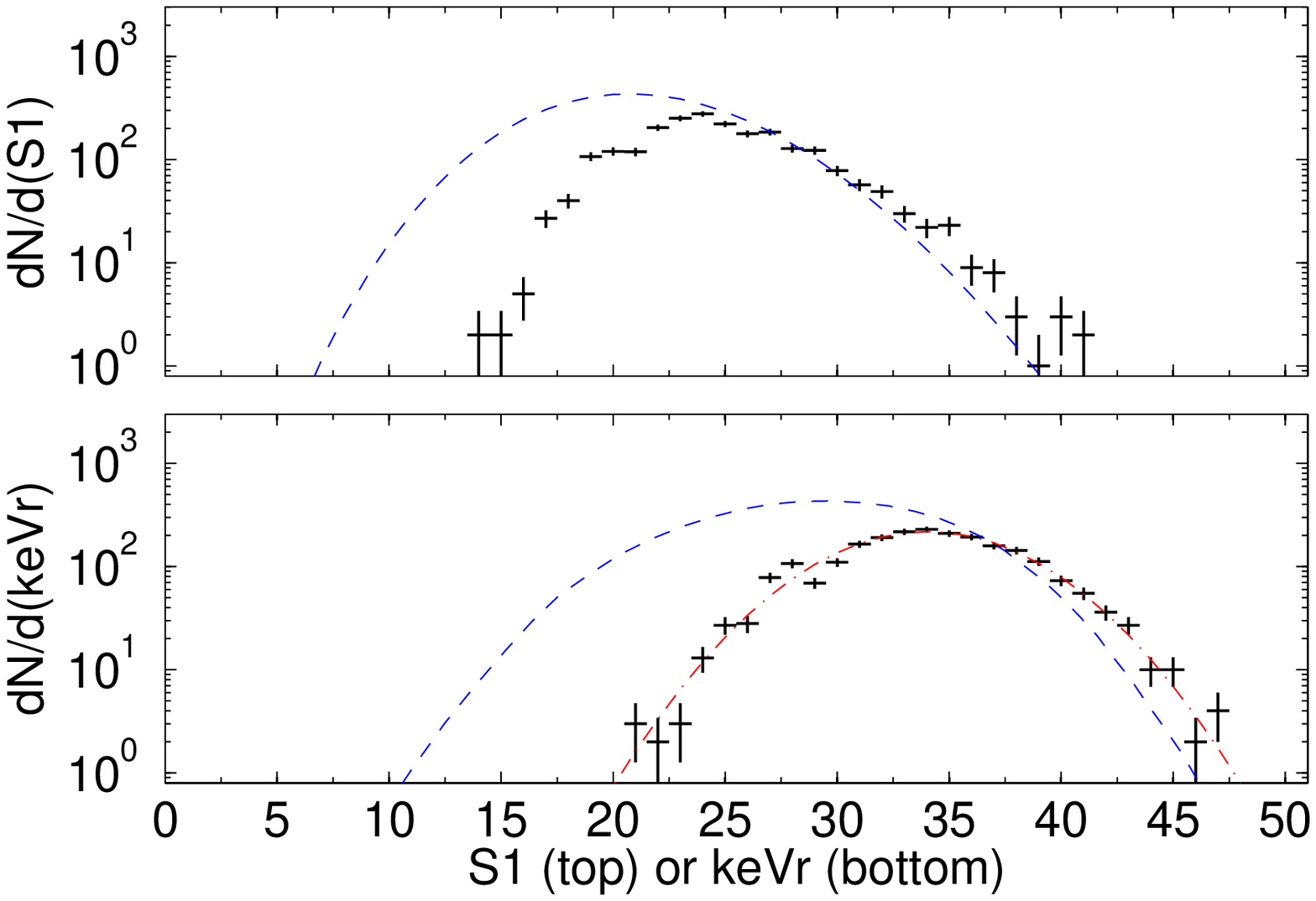,width=10cm}
\caption{All symbols and curves as described in Fig. \ref{fig8}. The Poisson distribution (blue dash) has $k=21$.}
\label{fig9}
}

A second example is shown in Fig. \ref{fig9}, for E$_{nr}=30$~keV.  The left edge of the distribution is now truncated solely by the top of the 50\% acceptance box.  The right hand edge is visibly widened, relative to the Poisson expectation (blue dash), by photomultiplier fluctuations.   In terms of keVr, a Gaussian fit (red dash-dot) provides a reasonable parameterization.  This is true for E$_{nr}\gtrsim25$~keV.

It seems unlikely that the detector resolution can be summarized with a simple statistical model, particularly at low recoil energies where it is most relevant.  For light mass ($\lesssim10$~GeV) particle dark matter, in which case detector sensitivity is dominated by Poisson fluctuations above the S1 threshold, a correct model of the detector resolution is essential.  As a practical matter, it seems preferable to use the discrete resolution distributions obtained directly from the simulation \cite{extramat}.

\subsection{Cut acceptance $\epsilon$}
The cut acceptances $\epsilon$ reported by e.g. the XENON10 collaboration \cite{2009angle,2008angle} or XENON100 collaboration \cite{2010aprile} as a function of detected S1 photoelectrons reflect the fraction of nuclear recoil events passing all cuts, in each S1 bin.  A correct mapping between S1 and keV requires considering the probabilities that a measured S1 value corresponded to each value of keV nuclear recoil energy.  In calculating dark matter exclusion limits, it is therefore more natural to scale the bin edges for each $\epsilon$ to keVr via $\mathcal{L}_{eff}$; they should then be applied to the theoretical prediction for the differential scattering rate (Eq. \ref{rate_eq}) only after correcting for detector resolution.

\section{XENON100} \label{xenon100}
The results obtained in Sec. \ref{nrbands} and Sec. \ref{lowedet} also apply to the XENON100 experiment, with slight numerical modification.  The major difference from a sensitivity point of view is the lower $L_y/S_e=2.2/0.58$ reported by the XENON100 collaboration \cite{2010aprile}.  This is the primary reason that XENON100 has a reduced sensitivity to low mass dark matter, relative to XENON10.  The superior sensitivity of XENON10 in the low mass range has already been noted \cite{2010savage}, but attributed solely to the lower S1 threshold (2~photoelectrons at XENON10, versus a stated 4~photoelectrons at XENON100).  XENON100 reports \cite{2010aprile} a slightly lower S2 gain of about 20~photoelectrons per extracted electron.  Practically, this appears to translate into a 15~electron (rather than 12~electron) S2 software threshold.  However, the S2 threshold has a minimal impact on $\eta$, since $\eta$ is still dominated by the S1 $n\ge2$ coincidence requirement and $L_y/S_e$.  
%In Sec. \ref{dmlimits} we show the effect of $L_y/S_e$, given equal S1 thresholds.

\FIGURE{
\epsfig{file=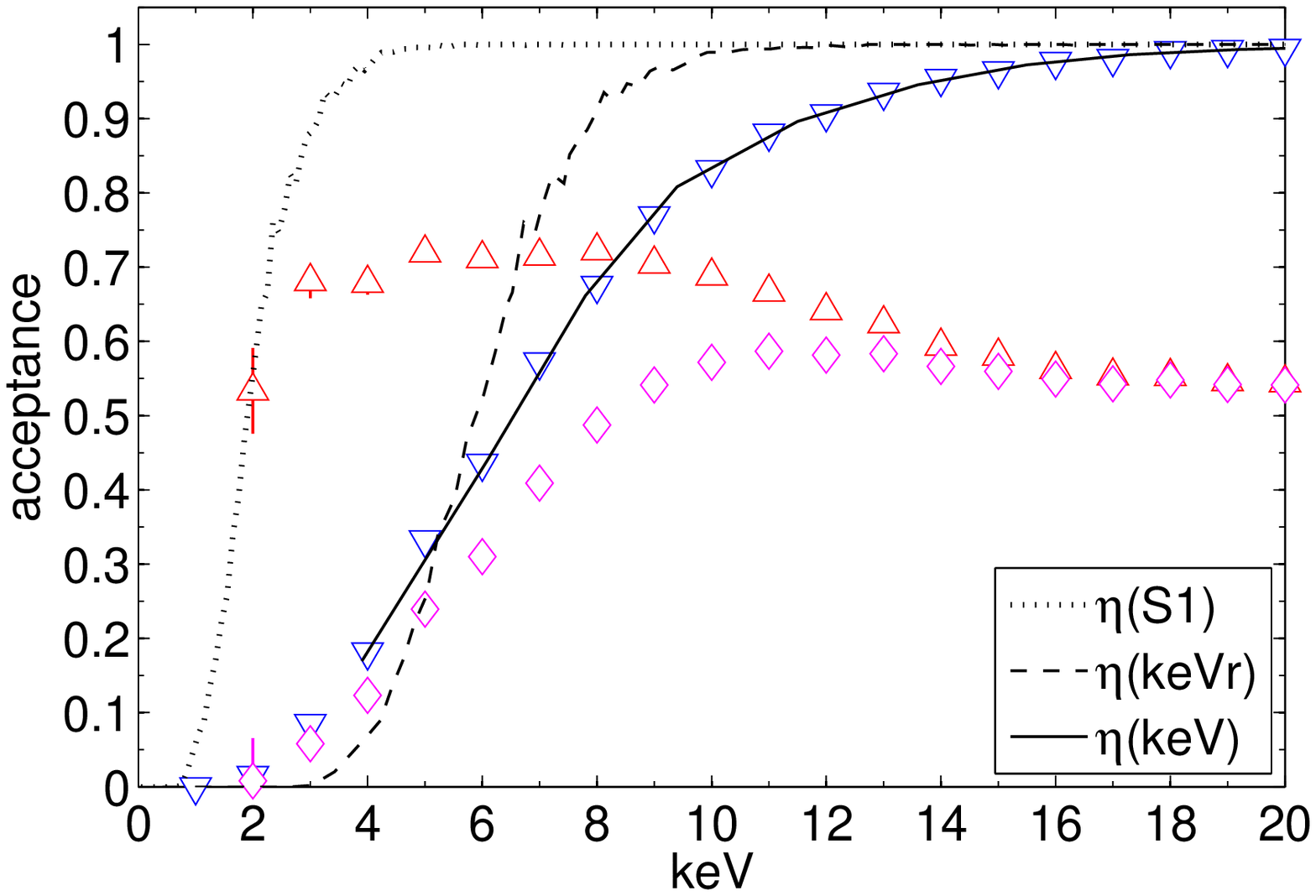,width=10cm}
\caption{XENON100 detector acceptance, assuming $\mathcal{Q}_y$ and $\mathcal{L}_{eff}$ given by {\it Case 2} from Fig. \ref{fig0}.  All symbols and curves are as described in Fig. \ref{fig7}.}
\label{fig10}
}

Finally, we note that the XENON100 collaboration have opted to require the S1 $n\geq2$ coincidence within a 20~ns time window \cite{2010aprile}.  There is some uncertainty in the scintillation decay time for nuclear recoils in liquid xenon, and in the Appendix we describe a reasonable parameterization.  With this parameterization, we find that a 20~ns coincidence window (rather than 300~ns) reduces $\eta$ by 20\% at 5~keV, and about 10\% at 10~keV.  In Fig. \ref{fig10} we show the detector efficiency obtained with the XENON100 detector parameters, given the physical assumptions of {\it Case 2} from Fig. \ref{fig0}.  Excellent agreement with the XENON10 nuclear recoil band was obtained (but is not shown).  This should not be surprising given the instrumental similarity between the two experiments.  We note that the XENON100 collaboration have already confirmed the similarity of the nuclear recoil bands \cite{2010aprile_wonder} obtained from the two experiments.  The only notable difference we observe is a slight upturn in the simulated XENON100 $\mu$ values, below about 15 S1 photoelectrons.  This was also observed in the XENON100 data \cite{2010aprile_wonder}.

\section{Dark matter exclusion limits} \label{dmlimits}
Having obtained significant insight on the low energy acceptance $\eta$, the actual acceptance of the 50\% acceptance box and the detector resolution, we turn to calculation of dark matter exclusion limits.  As in Sec. \ref{nrbands} and Sec. \ref{lowedet}, we use the $\mathcal{Q}_y$ and $\mathcal{L}_{eff}$ curves given by {\it Case 2} as a central value.  Noting that any reasonable $\mathcal{L}_{eff}$ curve must also lead to reasonable values for $\mathcal{Q}_y$, we choose the set of curves shown in Fig. \ref{fig00} (dashed) as a lower bound.  This is conservative considering that lower values of $\mathcal{Q}_y$ would be very difficult to reconcile with measurements \cite{2010manzur}.  We also consider a spline through the measured values of \cite{2009aprile} as an upper bound for $\mathcal{L}_{eff}$, as shown in Fig. \ref{fig00} (dotted).  While higher values of $\mathcal{L}_{eff}$ could also be considered reasonable, we find it more interesting to consider a case which coincides with a published measurement.  Each of these cases results in excellent agreement with the XENON10 nuclear recoil band, similar to that shown in Fig. \ref{fig2}.

\FIGURE{
\epsfig{file=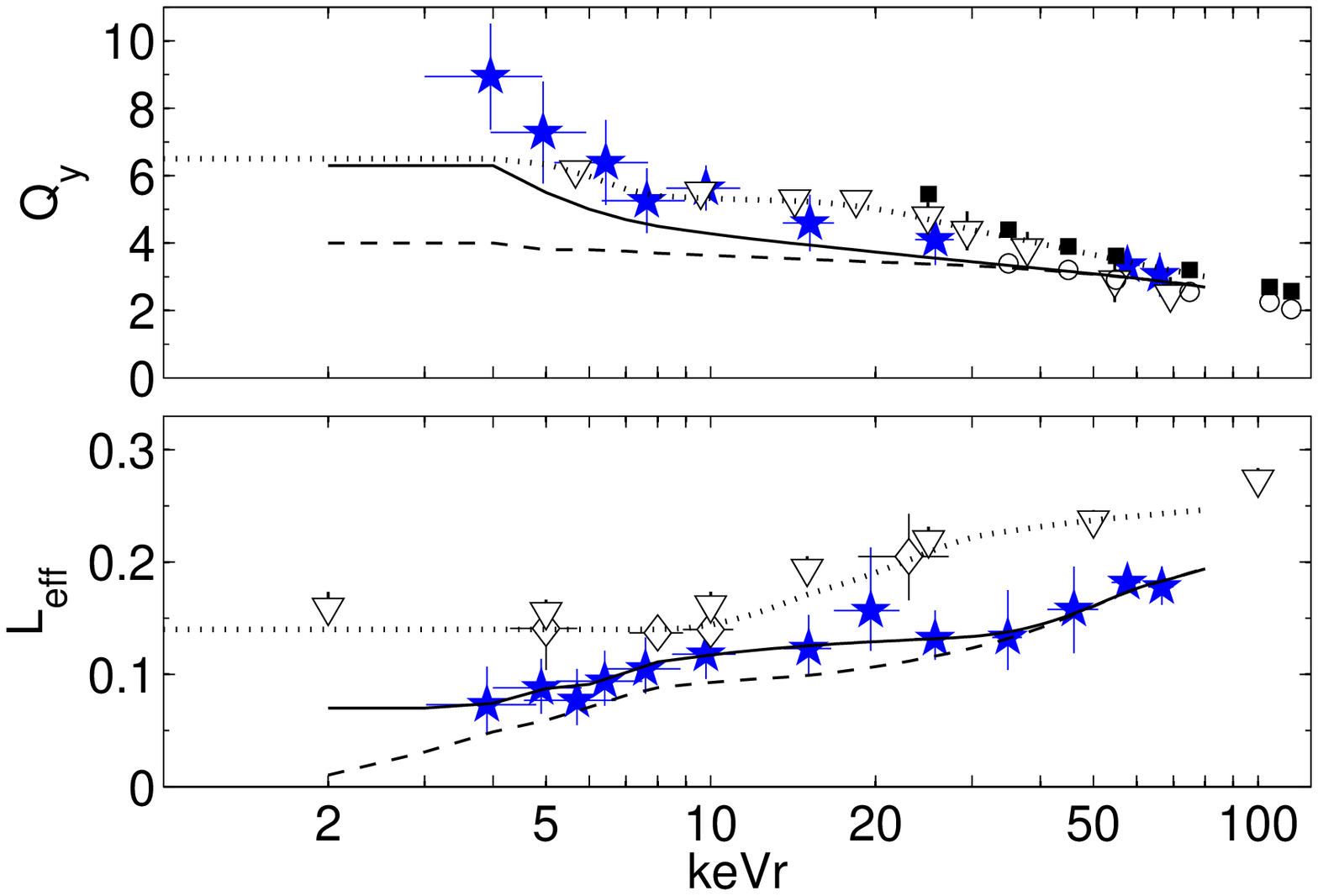,width=10cm}
\caption{$\mathcal{Q}_y$ and $\mathcal{L}_{eff}$ values used to calculate central (solid curve, also shown in Fig. \ref{fig0} as {\it Case 2}) most conservative (dashed) and upper (dotted) 90\% C.L. bounds.  Each curve is extrapolated as necessary to reproduce the measured nuclear recoil band shown in Fig. \ref{fig2}.  The dotted curve is truncated at 1~keV.  Other data as indicated in Fig. \ref{fig0}.}%Also shown are data from \cite{2010manzur} (blue $\bigstar$), \cite{2009aprile} ($\lozenge$), \cite{2009sorensen} ($\triangledown$) and \cite{2006aprile} ($\circ$ and $\blacksquare$).}
\label{fig00}
}

We calculate predicted differential event rates as a function of nuclear recoil energy on a xenon target from 
\begin{equation} \label{rate_eq}
\frac{dR}{d\mbox{E}_{nr}} = N_T M_N \frac{\rho_{\chi} \sigma_n}{2 m_{\chi} \mu_{ne}^2}
     A^2 %\frac{(f_pZ+f_n(A-Z))^2}{f_n^2} 
     F^2(\mbox{E}_{nr}) \int_{v_{min}}^{v_{esc}} \frac{f(v)}{v} dv,
\end{equation}
where the number of target nuclei in the detector is $N_T$, the mass of the target nucleus is $M_N$ and its atomic number is $A$.  We assume the standard local dark matter density $\rho_{\chi} = 0.3$~GeV~cm$^{-3}$, with dark matter particle mass $m_{\chi}$ and cross section (per nucleon) $\sigma_n$. The reduced mass $\mu_{ne}$ is for the nucleon $-$ dark matter particle system.  The nuclear form factor $F(\mbox{E}_{nr})$ accounts for a loss of coherence as momentum transfer to the nucleus increases.  We use the Helm form factor parameterization $F(\mbox{E}_{nr}) = 3j_1(qr_n)/qr_n \cdot \exp(-(qs)^2/2)$.  We take the effective nuclear radius $r_n = \sqrt{r_0^2-5s^2}$, with $r_0=1.2A^{1/3}$~fm and the skin thickness $s=1$~fm \cite{1996lewin}.  The momentum transfer to the nucleus is just $q=\sqrt{2M_N\mbox{E}_{nr}}$.

\FIGURE{
\epsfig{file=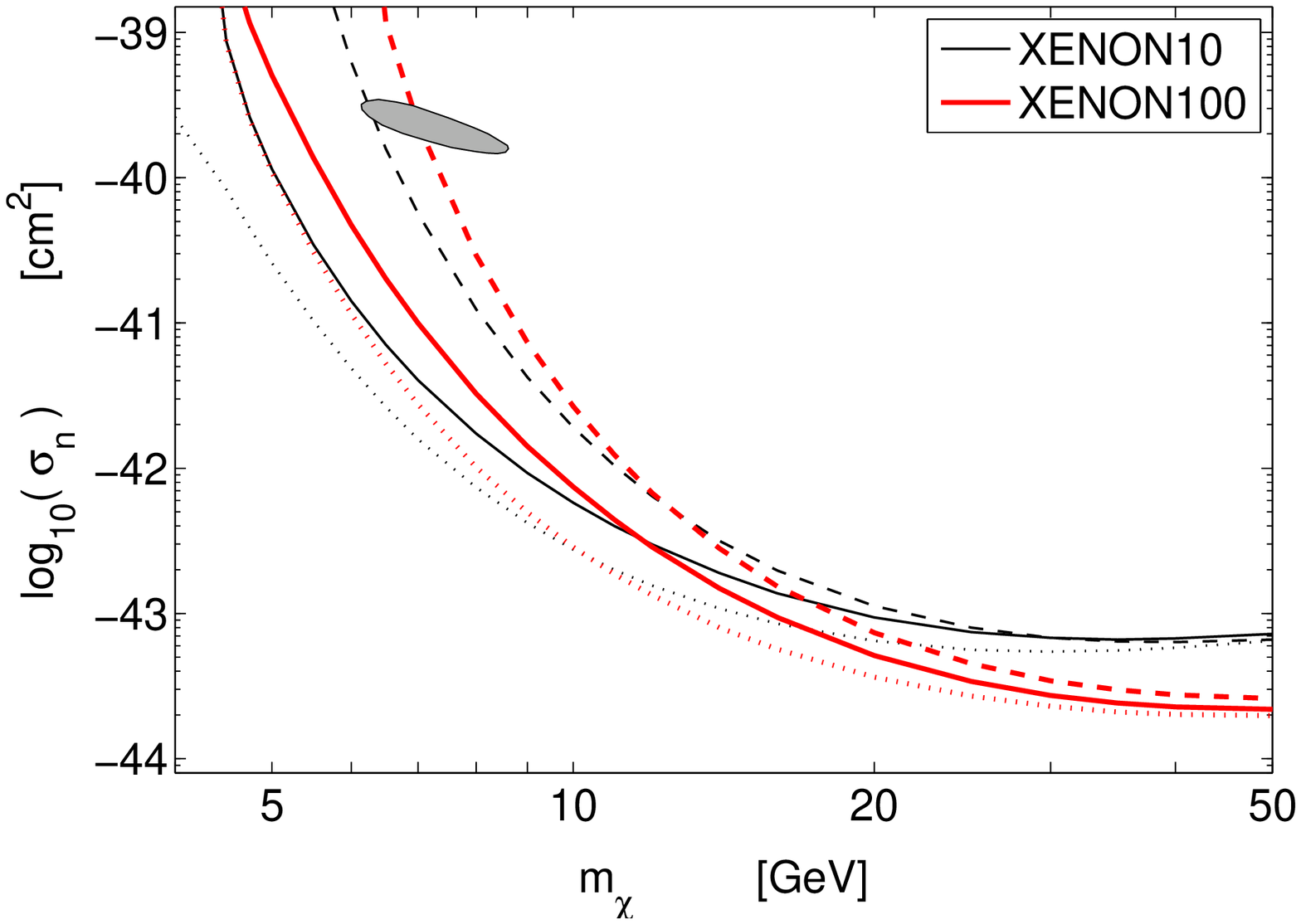,width=10cm}
\caption{For both XENON10 and XENON100 we show exclusion limits in the $m_{\chi}-\sigma_n$ plane, for each of the three $\mathcal{L}_{eff}$ curves shown in Fig. \ref{fig00} (with corresponding line style).  The shaded ellipse corresponds to a combined CoGeNT+DAMA 99\% allowed region \cite{2010hooper}.  Parameter space above the curves is excluded at 90\% C.L.}
\label{fig11}
}

We assume the dark matter halo velocity distribution to be Maxwellian, and perform the integration over $f(v)/v$ following \cite{2006savage}.  We take the velocity dispersion to be $v_0=230$~km~s$^{-1}$, and as in \cite{2010chang} set the rotational speed of the local standard of rest $v_{rot}=v_0$.  Considering the sun's velocity \cite{2006savage} implies an earth-halo velocity of 243~km~s$^{-1}$.   The lower limit of the integral in Eq. \ref{rate_eq} is $v_{min} = \sqrt{1/(2m_N\mbox{E}_{nr})}(m_N\mbox{E}_{nr}/\mu)$.  The upper limit is set by the galactic escape velocity, which we take to be $v_{esc}=600$~km~s$^{-1}$ \cite{2007smith}.

In calculating 90\% C.L. exclusion limits, we follow the procedure described in Sec. \ref{etakeV}.  The combined low-energy efficiency $\eta$ (Fig. \ref{fig7} and Fig. \ref{fig10}, $\lozenge$) is applied directly to the predicted nuclear recoil energy spectrum (Eq. \ref{rate_eq}).  The spectrum is then convolved with the detector resolution as obtained directly from the simulation \cite{extramat}.  The stated cut acceptances are then applied.  These are given for XENON10 by Table I of \cite{2009angle}, and XENON100 by Fig. 2 of \cite{2010aprile}. For XENON100, we presume their total acceptance curve \cite{2010aprile} already includes their estimate of $\eta$(S1).  As shown in Fig. \ref{fig7} , this should roll off rapidly below 4~photoelectrons.  Since we have already applied $\eta$ to Eq. \ref{rate_eq}, we take the stated XENON100 acceptance to be a constant 0.65 below 5~photoelectrons.  We also note that the XENON100 acceptance box is defined between $\mu$ and the 15~electron S2 software threshold \cite{2010aprile}.  This may result in a slightly lower acceptance at low energies, relative to an acceptance box defined between $\mu$ and $\mu-3\sigma$.  We assume such an effect, if present, is negligible.

The lower limit of integration for calculating the number of events predicted by Eq. \ref{rate_eq} is determined by the S1 threshold, converted to keVr via $\mathcal{L}_{eff}$.  For both experiments, we use 2~photoelectrons for this lower limit.  This is reasonable despite the stated 4~photoelectron analysis threshold of XENON100, as they observe no events above either threshold.  If we had instead used a 4~photoelectron S1 threshold for XENON100, those exclusion limit curves would have been about 30\% higher for $m_{\chi}\lesssim10$~GeV.  We reiterate that $\mathcal{L}_{eff}$ has only been extrapolated as shown in Fig. \ref{fig00}, which was necessary to reproduce the observed XENON10 nuclear recoil band.  We also note that because of the detector acceptance ($\eta$), any extrapolation below 2~keV is largely irrelevant.  

\section{Summary}
Our analysis has shown that reasonable bounds on the energy dependence of $\mathcal{L}_{eff}$ can be deduced from existing data on $\mathcal{Q}_y$, in conjunction with nuclear recoil band data.  These bounds, with their requisite extrapolation, are shown in Fig. \ref{fig00}.  This result makes use of the simple observation that the energy dependence of $N_e/N_{\gamma}$, and therefore also $\mathcal{Q}_y/\mathcal{L}_{eff}$, are encoded in the nuclear recoil band log$_{10}$(S2/S1).  It has already been pointed out that the sensitivity of liquid xenon detectors to light  dark matter is substantially enhanced by considering the finite resolution \cite{2010aprile}.  However, we have shown that the detector resolution cannot be correctly described by only considering Poisson fluctuations in the number of detected S1 photoelectrons.  This is due in large part to the effect of taking 50\% acceptance for nuclear recoils, as described in Sec. \ref{fiftybox}.  We have made available \cite{extramat} the discrete resolution distributions for each $\mathcal{L}_{eff}$ and $\mathcal{Q}_y$ case shown in Fig. \ref{fig00}.   We have also shown a rigorous treatment of liquid xenon detector acceptance, for low energy nuclear recoils.  The acceptance shown in Fig. \ref{fig7} and Fig. \ref{fig10} assume that position-dependence of signals leads to sub-dominant effects.  If this were not true, the actual detector acceptance could be lower.

For equal S1 thresholds of 2 photoelectrons, XENON10 shows more sensitivity than does XENON100 to dark matter with particle mass $m_{\chi}\lesssim30$~GeV.  This is a result of the difference in $\eta$ between the two experiments (compare Fig. \ref{fig7} with Fig. \ref{fig10}), which in turn is driven almost exclusively by the higher $L_y/S_e$ obtained by XENON10.  An S1 coincidence requirement of $n\geq3$, which was not employed by either experiment, would be a good example of another factor which that would significantly lower  $\eta$.  Our results indicate that at 90\% C.L. the XENON10 data are inconsistent with the combined CoGeNT+DAMA 99\% C.L. allowed region \cite{2010hooper}, given our central (Fig. \ref{fig00}, solid curve) assumption about $\mathcal{L}_{eff}$.  With the most conservative assumption about $\mathcal{L}_{eff}$ (Fig. \ref{fig00}, dashed curve), XENON10 (XENON100) is marginally (comfortably) consistent with the combined CoGeNT+DAMA 99\% C.L. allowed region.  Taken individually, either the CoGeNT or DAMA allowed regions \cite{2010hooper} remain consistent with the XENON10 constraints;  these cases are not shown in Fig. \ref{fig11}.  A separate analysis of the CoGeNT data in conjunction with a 50\% exponential background contribution finds consistent parameter space for $m_{\chi}\lesssim6$~GeV \cite{2010chang}.  We also note that our exclusion limits appear to be less stringent than those derived in \cite{2010savage}, for $m_{\chi}\lesssim6$~GeV.  This is probably a result of the treatment of $\eta$ and the detector resolution.  

We note that neither astrophysical uncertainties nor uncertainty in the form factor $F(\mbox{E}_{nr})$ have been taken into account in the exclusion limits.  Considering these factors will likely ease the tension between the null results of XENON10 and XENON100, and the dark matter interpretation of CoGeNT and DAMA.  We also remind the reader that the consistency of {\it Case 2} described in Sec. \ref{case2} required an approximately $1\sigma$ systematic decrease in the measured value of $\mathcal{Q}_y$ from \cite{2010manzur}.  This is consistent with expectations, and it highlights the need to obtain both $\mathcal{Q}_y$ and nuclear recoil band data under exactly the same operating conditions.  We do not expect the acquisition of such data to significantly modify our conclusions, but rather to decrease the range of reasonable values of the scintillation yield from that shown in Fig. \ref{fig00}.

\acknowledgments{This work benefitted from many discussions during the Aspen Center for Physics workshop {\it ``From Colliders to the Dark Sector: Understanding Dark Matter at Particle Colliders and Beyond,''} Aspen CO, June 2010.  I would like to thank the workshop organizers, and the ACP for their hospitality during that time.  I gratefully acknowledge discussions with Adam Bernstein, Rick Gaitskell, Aaron Manalaysay, Dan McKinsey and Neal Weiner.  I also thank Thomas Schwetz for pointing out an error in an early version of Fig. \ref{fig8}.}

\appendix
\section{Scintillation decay time constant $\tau$}
In this Appendix we briefly discuss the scintillation decay time constant ($\tau$) for liquid xenon.  The relevance of this discussion centers on the choice of the XENON100 collaboration to require $n\ge2$ coincidence in a 20~ns time window.  For the 300~ns coincidence window used by XENON10 it is largely irrelevant, since $>99.9\%$ of scintillation occurs within 300~ns in any case.  It is well established that the excited molecular state Xe$_2^*$ exists in either a triplet or singlet spin configuration, with a different relaxation time in each case \cite{1978kubota}.  In other words, the primary scintillation signal as a function of time is
\begin{equation} \label{scint}
S1(t) = A_s \cdot e^{(-t/\tau_s)} + A_t \cdot e^{(-t/\tau_t)}.
\end{equation}
Measurements of $\tau_s=2$~ns and $\tau_t=27$~ns for electron recoils, and $\tau_s=4$~ns and $\tau_t=22$~ns for nuclear recoils can be found in \cite{1999doke}.  We are not aware of any physical reason why the singlet and triplet state decay times should depend on the incident particle species, and so assume $\tau_s=4$~ns and $\tau_t=27$~ns.

\FIGURE{
\epsfig{file=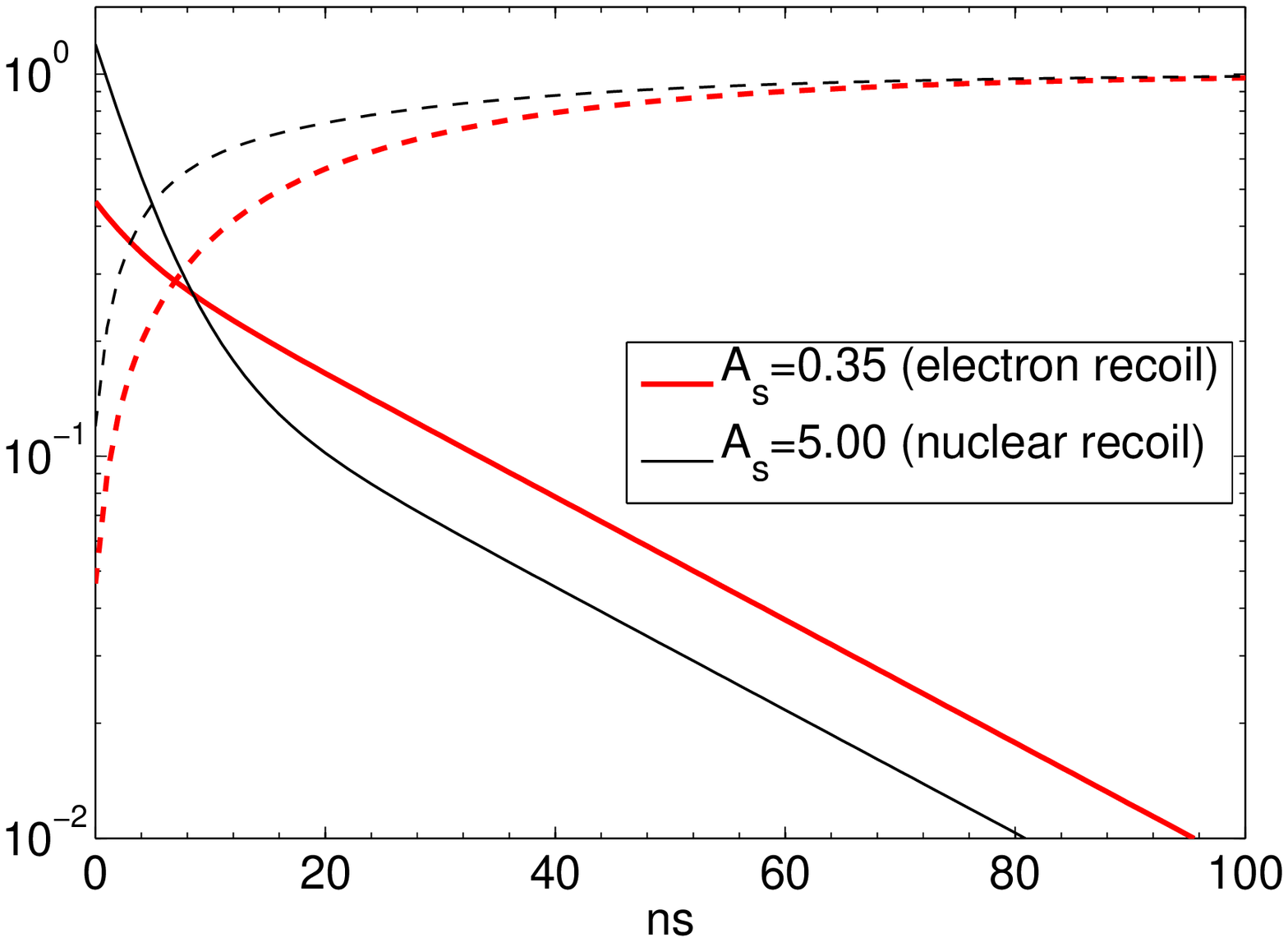,width=10cm}
\caption{Idealized primary scintillation pulse shapes (solid curves) for nuclear ($A_s=5.00$) and electronic ($A_s=0.35$) recoils.  In both case, $A_t=1$.  The cumulative sum of each pulse is also indicated (dashed curves).  Note that the pulse shapes, which are normalized to unit area, have been scaled by $\times10$ for display.}
\label{figA}
}

The intensity ratio $I_s/I_t\equiv A_s\tau_s/A_t\tau_t$, is often used to indicate the relative strength of each component.  For electron recoils, $I_s/I_t=0.05$ \cite{1978kubota} and for recoils resulting from alpha particles, $I_s/I_t=0.43$ \cite{1999doke}.  We are not aware of any measurement for low-energy nuclear recoils from neutrons, but we can infer the intensity ratio from the prompt fraction of scintillation reported by the XENON10 collaboration (Fig. 2 of \cite{2009angle}).  We calculate the prompt fraction $f_p$ as defined in \cite{2009angle,2009kwong} for the idealized pulse defined by Eq. \ref{scint}, with no attempt to account for pulse shaping due to electronics.  We choose $A_s=0.35$ and $A_t=1$ for electron recoils, since these values reproduce the measured $I_s/I_t=0.05$.  This leads to $f_p=0.19$, in reasonable agreement with \cite{2009angle}.  Noting that \cite{2009angle} report an asymptotic value of $f_p$ for nuclear recoils that is about 12\% larger than their electron recoil value, we then ask what value of $A_s$ this implies for nuclear recoils.  It turns out that $A_s=5.00$ and $A_t=1$ ($I_s/I_t=0.74$) provide consistency with that work.

This implies that about 75\% of scintillation photons from nuclear recoils would be expected to appear in the first 20~ns, as summarized in Fig. \ref{figA}.  Note that if it were the case that $A_t=0$ for nuclear recoils, 99\% of scintillation photons would be expected in the first 20~ns.  However, the data \cite{1978kubota,1999doke} do not support such a conjecture, nor does the prompt fraction reported in \cite{2009angle,2009kwong}.  Given these concerns, a direct measurement of the intensity ratio for nuclear recoils is probably in order.  Whether or not that occurs, the XENON100 collaboration may wish to increase their coincidence time window to e.g. $100$~ns, for a more comfortable $99\%$ collection of the S1 signal.

\clearpage{\pagestyle{empty}\cleardoublepage}

\end{document}